\documentclass[prl,twocolumn,showpacs,preprintnumbers,amsmath,amssymb,superscriptaddress]{revtex4-1}

\usepackage{graphicx}% Include figure files
\usepackage{epsfig}
\usepackage{amsmath}
\usepackage{textcomp}
\usepackage{amssymb}
\usepackage{dcolumn}% Align table columns on decimal point
\usepackage{bm}% bold math
\usepackage{braket}
\PassOptionsToPackage{numbers,sort&compress}{natbib}

\makeatletter
\g@addto@macro\normalsize{%
  \setlength\abovedisplayskip{4pt}
  \setlength\belowdisplayskip{4pt}
  \setlength\abovedisplayshortskip{4pt}
  \setlength\belowdisplayshortskip{4pt}
}
\makeatother

\newcommand{\be}{\begin{equation}}
\newcommand{\ee}{\end{equation}}
\newcommand{\bea}{\begin{eqnarray}}
\newcommand{\eea}{\end{eqnarray}}

\begin{document}

\title{Fast ion swapping for quantum information processing}

\author{H.~Kaufmann}
\author{T.~Ruster}
\author{C.~T.~Schmiegelow}\thanks{Present address: LIAF -  Laboratorio de Iones y Atomos Frios, Departamento de Fisica \& Instituto de Fisica de Buenos Aires, 1428 Buenos Aires, Argentina}
\author{M.~A.~Luda}\thanks{Present address: DEILAP, CITEDEF \& CONICET, J.B. de La Salle 4397, 1603 Villa Martelli, Buenos Aires, Argentina}
\author{V.~Kaushal}
\author{J.~Schulz}
\author{D.~von Lindenfels}
\author{F.~Schmidt-Kaler}
\author{U.~G.~Poschinger}\email{poschin@uni-mainz.de}

\affiliation{Institut f\"ur Physik, Universit\"at Mainz, Staudingerweg 7, 55128 Mainz, Germany}

\begin{abstract}
We demonstrate a SWAP gate between laser-cooled ions in a segmented microtrap via fast physical swapping of the ion positions. This operation is used in conjunction with qubit initialization, manipulation and readout, and with other types of shuttling operations such as linear transport and crystal separation and merging. Combining these operations, we perform quantum process tomography of the SWAP gate, obtaining a mean process fidelity of 99.5(5)\%. The swap operation is demonstrated with motional excitations below 0.05(1)~quanta for all six collective modes of a two-ion crystal, for a process duration of 42~$\mu$s. Extending these techniques to three ions, we reverse the order of a three-ion crystal and reconstruct the truth table for this operation, resulting in a mean process fidelity of 99.96(13)\% in the logical basis.

\end{abstract}


\maketitle

%\section{Introduction}
The last decade has seen substantial progress towards scalable quantum computing with trapped ions. Gate fidelities reach fault-tolerance thresholds \cite{BALLANCE2015}, and first steps towards realizing decoherence-free qubits have been demonstrated \cite{NIGG2014}. Moreover, microfabricated, segmented ion traps continue to mature as an experimental low-noise environment \cite{HITE2012,DANIILIDIS2014} hosting multi-qubit systems \cite{HOME2009,HEROLD2015}. In the seminal proposal from Kielpinsky, Monroe and Wineland \cite{KIELPINSKI2002} for such \textit{quantum CCD chip}, scalability is reached through ion shuttling operations, where trapped-ion qubits are moved between different trap sites through application of suitable voltage waveforms to the trap electrodes. Since the first demonstration of ion shuttling in segmented traps \cite{ROWE2002}, the development of trap control hardware has progressed \cite{BOWLER2013,BAIG2013}. This has recently led to demonstrations of fast ion shuttling at low final motional excitation \cite{WALTHER2012,BOWLER2012}.\\
It is currently an open question if a trapped-ion quantum computer should be based on large processing units hosting thousands of qubits \cite{STEANE2007,LEKITSCH2015} or on a modular architecture of medium-sized nodes with photonic interconnectivity \cite{MONROE2014}. With current technology, the possibilities for high-fidelity coherent control and readout of ion strings consisting of more than a few ions are limited, such that ion shuttling is required in either case. For universal quantum computation, two-qubit gates need to be performed between arbitrary pairs of ions, such that reordering ion strings becomes a necessary. Furthermore, if multiple ion species \cite{HOME2013} are employed for sympathetic cooling \cite{KIELPINSKI2000} or ancilla-based syndrome readout via inter-species entangling gates \cite{BALLANCE2015HYBRID,TAN2015}, deterministic ion reconfiguration is ultimately required.\\

\begin{figure}[h!tp]\begin{center}
\includegraphics[width=0.40\textwidth]{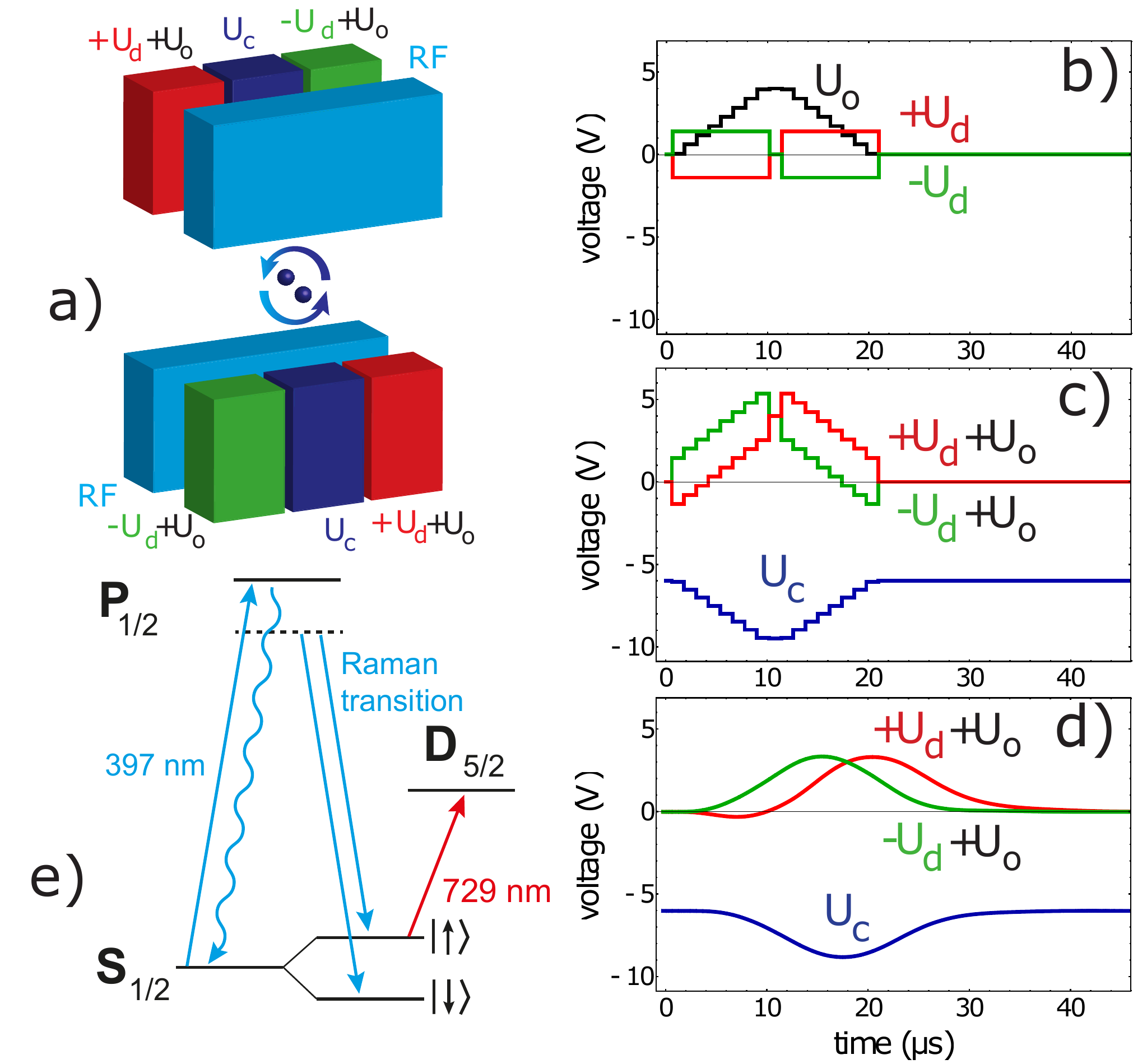}
\caption{Ion swapping in a multilayer segmented trap. \textbf{a)} shows the relevant trap electrodes, indicating where trapping voltage $U_c$, the diagonal voltage $U_d$ and the offset voltage $U_o$ controlling the process are applied. Panels \textbf{b)} and \textbf{c)} show the voltage ramps in the form of discrete samples, as they are programmed to the arbitrary waveform generator. Here b) shows $U_d$ and $U_o$, while c) shows the actual electrode voltages. Panel \textbf{d)} shows the voltage ramps, measured after the external 50~kHz second-order low-pass filter, which leads to smoothing, but also to delay and increase of duration of the ramps. \textbf{e)} shows the relevant part of the level scheme of $^{40}$Ca$^+$.
}
 \label{fig:swap}
\end{center}\end{figure}

To that end, segmented ion traps bearing junctions with T\cite{HENSINGER2006}, X\cite{BLAKESTAD2009,WRIGHT2013} or Y\cite{SHU2014} geometry have been developed and tested. Junctions increase the design complexity of the traps and allow only for sequential ion transport. Shuttling through junctions may yield large motional excitations, which precludes the execution of two-qubit gates. In this work, we perform ion reordering via on-site swapping of ions through application of suitable electric potentials. The advantages of this operation are that it does not require sophisticated electrode structures, and that parallel multi-site swappings could be performed.
While it has been shown \cite{SPLATT2009} that segmented traps allow for deterministic ion swapping, we demonstrate this process on fast timescales, comparable to qubit operation times. Importantly, motional excitation is avoided, such that the ions stay within the Lamb-Dicke regime for all six collective modes of vibration of a two-ion crystal. To highlight that this operation is deterministic and that it can be used in conjunction with other qubit operations, we integrate it within a sequence of shuttling, separation \cite{BOWLER2012,RUSTER2014,KAUFMANN2014} and merging operations and qubit manipulations to realize a full quantum process tomography of the SWAP gate. By performing the swap operation on near-ground-state cooled ions and combining it with qubit manipulations and other shuttling operations, we demonstrate its potential use for scalable quantum logic with trapped ions.

%%%%%%%%%%%%%%%%%%%%%%%%%%%%%%%%%%%%%%%%%%%%%%%%%%%%%%%%%%%%%%%%%%%%
%\section{Setup}
%%%%%%%%%%%%%%%%%%%%%%%%%%%%%%%%%%%%%%%%%%%%%%%%%%%%%%%%%%%%%%%%%%%%

\begin{figure}[h!tp]\begin{center}
\includegraphics[width=0.25\textwidth]{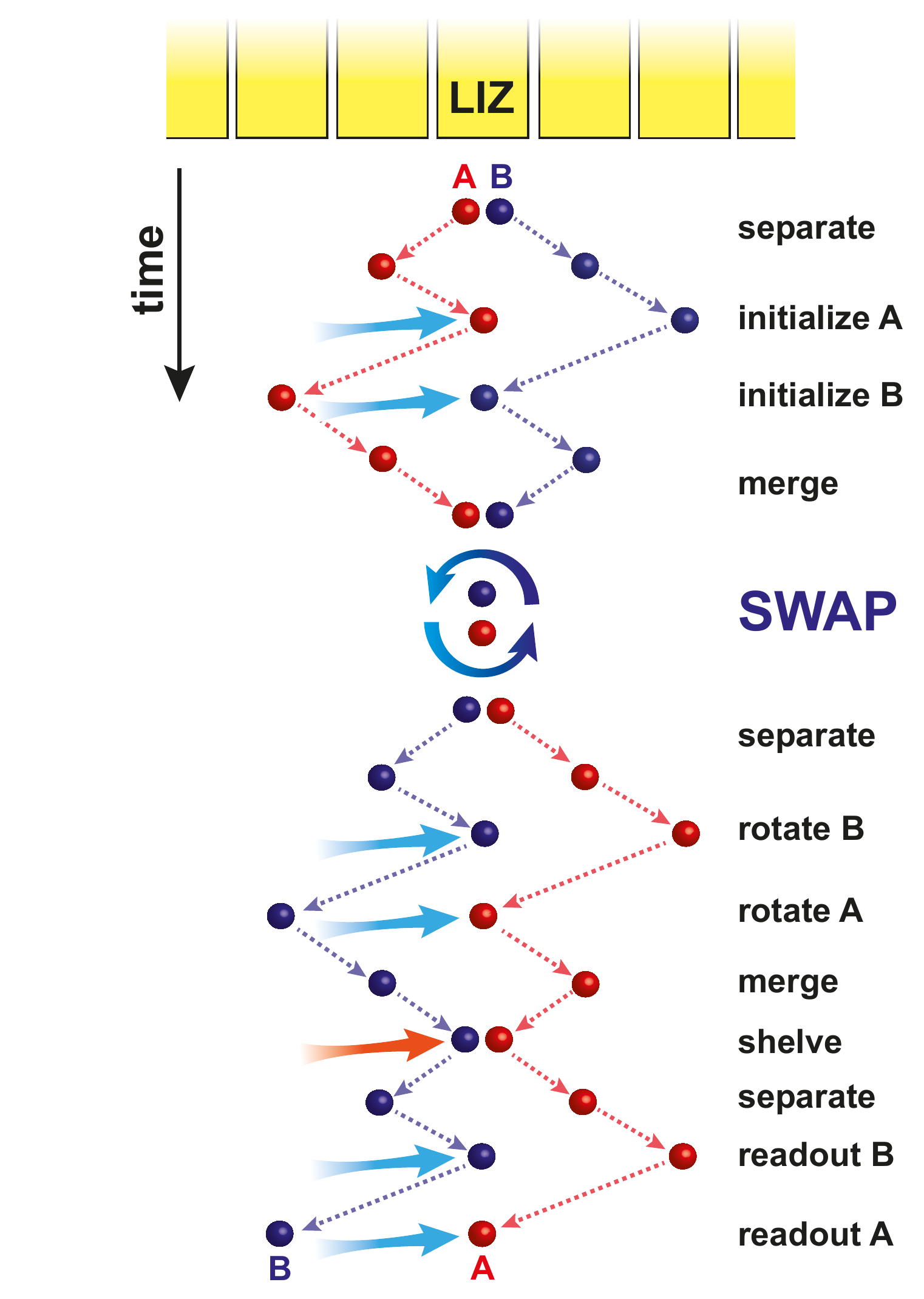}
\caption{Experimental quantum process tomography sequence for the SWAP operation. Each of the ion qubits A and B
is shuttled to the laser interaction zone for initialization laser pulses, followed by a fast SWAP operation and subsequent individual qubit rotation laser pulses. Finally the state is read out via electron shelving and fluorescence detection.
}
 \label{fig:sequence}
\end{center}\end{figure}

For our experiments, we trap $^{40}$Ca$^+$ ions in a segmented Paul trap similar to the design from \cite{SCHULZ2008}. Qubits are encoded in the Zeeman sublevels of the ground state $\ket{\downarrow}\equiv\ket{S_{1/2},m_J=-\tfrac{1}{2}}$ and $\ket{\uparrow}\equiv\ket{S_{1/2},m_J=+\tfrac{1}{2}}$, where an external magnetic field lifts the degeneracy by $2\pi\times$10.4~MHz. The ions are laser cooled on the S$_{1/2}\leftrightarrow$P$_{1/2}$ (cycling) dipole transition near 397~nm. Qubit initialization with a fidelity $>$99.8\% is achieved via optical pumping utilizing the narrow S$_{1/2}\leftrightarrow$D$_{5/2}$ quadrupole transition near 729~nm \cite{POSCHINGER2009}. For qubit manipulation, we employ stimulated Raman transitions mediated by a co-propagating pair of laser beams near 397~nm, detuned by $2\pi\times$250~GHz from the cycling transition. For sideband cooling and measurements of the motional state, we employ pairwise orthogonal propagating Raman beams, where the difference wavevector is aligned parallel (orthogonal) to the trap axis, providing coupling to axial (radial) modes of oscillation. Qubit readout is accomplished by spin-selective population transfer to the metastable $D_{5/2}$ state, followed by detection of state-dependent resonance fluorescence with a photomultiplier tube\cite{POSCHINGER2009}. All lasers are directed at the laser interaction zone (LIZ), see Fig. \ref{fig:sequence}. As the Raman beams driving the single-qubit operations are co-propagating, and the laser near 729~nm for electron shelving is directed perpendicularly to the trap axis, both operations are insensitive to ion motion along the trap axis. Control of the ion motion is achieved by individual supply of the trap electrodes by a fast multichannel arbitrary waveform generator \cite{WALTHER2012,RUSTER2014} at analog update rates of up to 2.5~MSamples/s, where each signal line has a second-order $\Pi$-type low-pass filter with a cutoff frequency of $2\pi\times$50~kHz.

%%%%%%%%%%%%%%%%%%%%%%%%%%%%%%%%%%%%%%%%%%%%%%%%%%%%%%%%%%%%%%%%%%%%
%\section{Shuttling  Swapping}
%%%%%%%%%%%%%%%%%%%%%%%%%%%%%%%%%%%%%%%%%%%%%%%%%%%%%%%%%%%%%%%%%%%%

\begin{figure}[h!tp]\begin{center}
\includegraphics[width=0.4\textwidth]{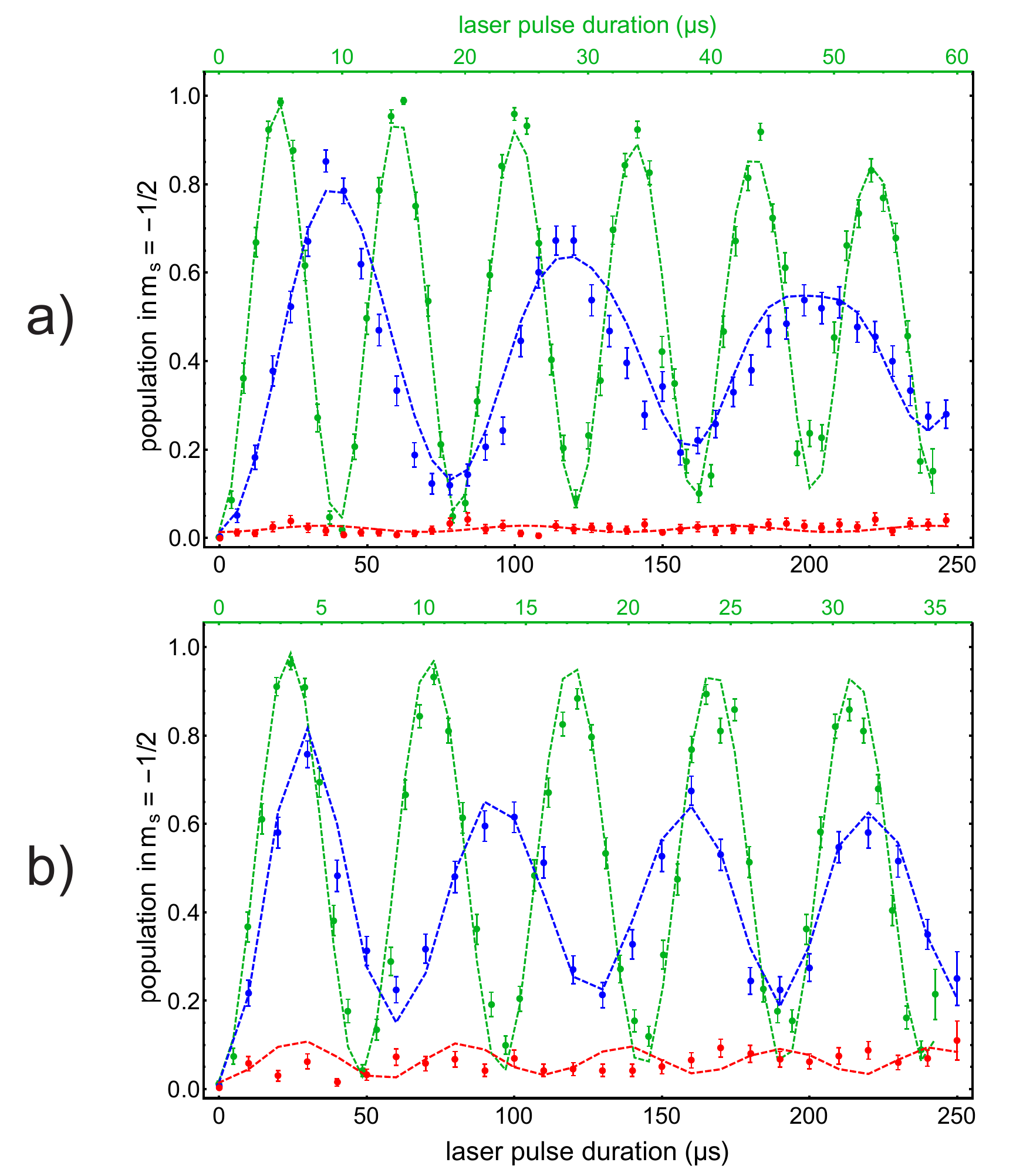}
\caption{Rabi oscillation data probed after swapping for verification of low-excitation swap operations: \textbf{a)} shows data for the axial center-of-mass mode, while \textbf{b)} shows data for the lower-frequency radial rocking mode. In both panels, the blue (red) points correspond to the blue (red) motional sideband, while the green points correspond to the carrier transition. The dashed lines are fits using model assuming oscillatory excitation. All curves indicate the probability for having at least one of the ions' state flipped to $\ket{\downarrow}$. Note that the time axis for the carrier data is scaled differently (upper axis ticks). Each data point corresponds to 200 state interrogations.
}
 \label{fig:excitation}
\end{center}\end{figure}

The on-site swapping process of a two-ion crystal is depicted in Fig. \ref{fig:swap}. We start with the crystal axially confined via a trapping voltage $U_c$ applied to the trapping electrode pair (blue segments in Fig. \ref{fig:swap}). Control over the rotation of the ion crystal is achieved via a diagonal, symmetry breaking dc quadrupole potential. This is generated by ramping up a voltage $+(-)U_d$ to the electrodes neighboring the trap site, shown in green (red) in Fig. \ref{fig:swap}. Here, the polarity on the electrode pair left of the trapping site is inverted as compared to the polarity of the electrode pair to the right. With the diagonal potential applied, the trapping voltage $U_c$ is gradually decreased, and a positive offset voltage $U_o$ is ramped up at the neighboring segments. The corresponding increase of the axial confinement drives the ion crystal through a structural transition from horizontal to vertical alignment. Simultaneously, the diagonal potential generated by $U_d$ is ramped down to 0~V. With the ion crystal vertically aligned, the process is conducted in reverse order, with inverted polarity of the voltage $U_d$ generating the diagonal potential.\\
We optimize the voltage ramps for the swap process by probing the final motional excitation on collective vibrational modes which are most affected, i.e. the axial \textit{stretch} and the lower frequency radial \textit{rocking} mode. The motional excitation is measured by driving Rabi oscillations on the red and blue motional sideband pertaining to all four radial and two axial collective modes of vibration. Each mode is cooled close to the ground state via resolved sideband cooling before the swapping operation, reaching mean phonon numbers between 0.016(4) (axial stretch mode) and 0.37(1) (lower frequency radial COM mode). Rabi oscillations are recorded over pulse areas in the range between $6\pi$ and $8\pi$ pertaining to the blue sideband Rabi frequency in the ground state. Fits assuming an oscillatory excitation, i.e. a coherent state of the corresponding mode, reveal the mean phonon number \cite{LEIBFRIED2003,WALTHER2012,RUSTER2014,SUPPLEMENTAL}.\\
Initially, a trapping voltage $U_c=$-6~V yields horizontal crystal alignment at an axial center-of-mass (COM) vibrational frequency of $2\pi\times$~1.488~MHz. The radial COM mode frequencies are $2\pi\times$~1.927~MHz and $2\pi\times$~3.248~MHz. We defining the dimensionless time $\tau=t/T$ for the total swapping time $T$. The least motional excitation is found for the following ramp parameters: The diagonal voltage $U_d$ is ramped up rapidly within $\tau=$0.05, to an optimum value of 1.4~V. For driving the ion crystal into vertical alignment, the axial COM frequency has to exceed the lower radial COM frequency. To that end, $U_c$ is ramped down to -9.5~V, while at the same time an additional offset voltage $U_o=$~+4~V is ramped up at all neighboring electrodes. Both $U_c$ and $U_o$ are ramped within $\tau=$0.05 to $\tau=$0.45. The polarity change of the diagonal voltage $U_d$ happens during $\tau=$0.45 to $\tau=$0.55. The resulting voltage ramps are depicted in Fig. \ref{fig:swap}. \\
The swapping operation was tested for increasing times $T$, until we found the shortest time with negligible motional excitation of $T=$22~$\mu$s, which -including the filters- corresponds to an actual duration of 42~$\mu$s. We measure the mean phonon number increase for all modes, comparing to the reference measurements directly after sideband cooling. For the axial modes, we find mean phonon number increases of 0.05(1)~phonons on the COM mode and 0.013(6)~phonons on the stretch mode. For the lower-frequency radial modes, corresponding to the plane in which the crystal rotates, we obtain 0.03(2)~phonons on the COM mode and 0.04(2)~phonons on the rocking mode. The higher-frequency radial mode, which is least affected from from the swapping, features 0.02(1)~phonons on the COM and 0.01(1)~phonons on the rocking mode. Rabi oscillation data probed after swapping is shown in Fig. \ref{fig:excitation} for the axial COM and lower frequency radial rocking modes.\\

Linear transport of ions along the trap axis is performed by gradually reducing the negative dc trapping voltage of $U_c=$~-6~V at the initial segment to 0~V, while applying a trapping voltage at the neighboring destination segment. We perform adiabatic transport at a duration of 28~$\mu$s per trap segment pair, spaced by 200~$\mu$m. Transport over more than one segment pair is performed by concatenation of this operation. Separation and merging operations require the transformation between single- and double-well potentials. The transient low axial confinement causes heating and oscillatory excitation \cite{KAUFMANN2014}. We employ tailored voltage waveforms and proper cancellation of residual forces along the trap axis, enabling separation and merging of two-ion crystals within 100~$\mu$s at a motional excitation of 5(2) quanta per ion \cite{RUSTER2014}. \\

%%%%%%%%%%%%%%%%%%%%%%%%%%%%%%%%%%%%%%%%%%%%%%%%%%%%%%%%%%%%%%%%%%%%
%\section{Process tomography}
%%%%%%%%%%%%%%%%%%%%%%%%%%%%%%%%%%%%%%%%%%%%%%%%%%%%%%%%%%%%%%%%%%%%

\begin{figure}[h!tp]\begin{center}
\includegraphics[width=0.5\textwidth]{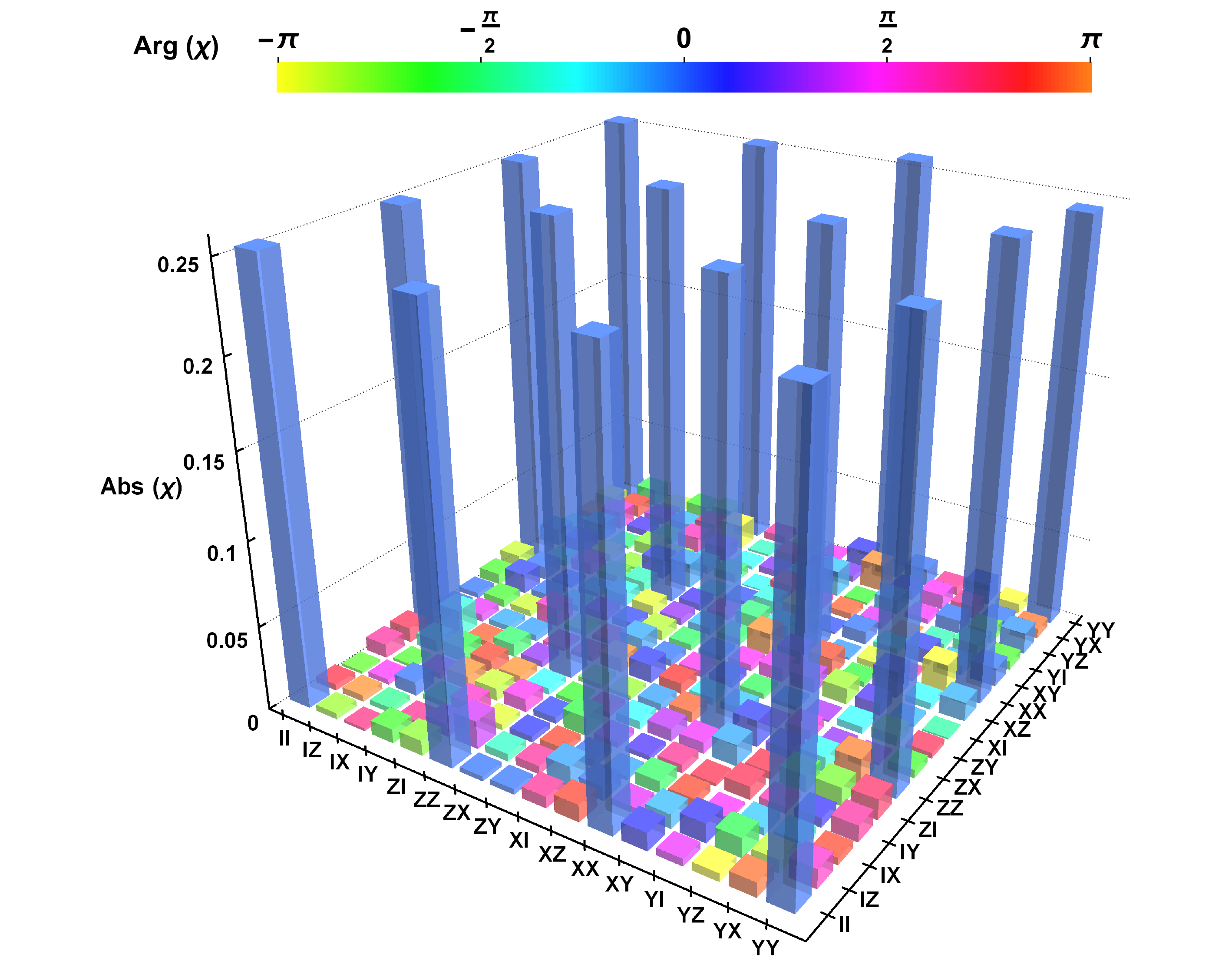}
\caption{Reconstructed process $\chi$-matrix for the SWAP operation. The absolute value of each matrix element is represented by the bar height, the phase is indicated by the color. The 16 elements which match the ideal absolute value of 0.25 have a controlled phase of $\arg(\chi_{ij})\approx 0$. All other elements are close to the ideal value of zero and have random phases. Each of the 144 preparation/measurement settings is probed on average 1260 times.
}
 \label{fig:tomo2}
\end{center}\end{figure}

The sequence for the process tomography consists of several shuttling operations and qubit manipulations, as depicted in Fig. \ref{fig:sequence}. First, the two-ion crystal is prepared by Doppler cooling and pumping in the LIZ. After separation, each qubit is individually shuttled into the LIZ, where one of the operations $\{\mathbb{1},R_X(\pi/2),R_Y(\pi/2),R_X(\pi)\}$ is applied to bring the respective qubit to the state $\{\ket{\uparrow},\ket{\uparrow}-i\ket{\downarrow},\ket{\uparrow}-\ket{\downarrow},\ket{\downarrow}\}$. The ions are recombined in the LIZ, where the swapping takes place. The crystal is again separated, and the ions are individually exposed to the analysis pulses $\{\mathbb{1},R_X(\pi/2),R_Y(\pi/2)\}$ for measuring the operators $\{\sigma_z,\sigma_y,\sigma_x\}$. After another merging operation in the LIZ, the population transfer $\ket{\uparrow} \leftrightarrow \ket{D_{5/2}}$ takes place. The ions are again separated and individually shuttled to the LIZ, where state dependent fluorescence is observed. Both qubits are shelved before fluorescence detection, to avoid depolarization of a remotely stored qubit from scattered light near 397~nm during the readout of the other qubit.
The analysis laser pulses have to be corrected for phases accumulated from moving the ions in the inhomogeneous magnetic field along the trap axis. Starting with the preparation pulse, qubit $i$ located at axial position $x_i(t)$ at time $t$ accumulates a phase which is determined by the deviation of the magnetic field from its value at the LIZ, $\Delta B(x)$, as 
\begin{equation}
\phi_i=\frac{\mu_B g_J}{\hbar} \int_{t_i^{(i)}}^{t_i^{(a)}} \Delta B(x_i(t)) dt.
\end{equation}
Here, $t_i^{(i)}$ denotes the instant of the state preparation pulse for qubit $i$ and $t_i^{(a)}$ denotes the instant for its analysis pulse. The magnetic field inhomogeneity along the trap axis is mapped out by using a single ion as a probe: Initialized in a superposition state, it is shuttled to the destination site $x$ and kept on hold for variable time $t$. After shuttling back to the LIZ, a refocusing $\pi$-pulse is applied, followed by another wait time of duration $t$ with the ion placed at the LIZ. Finally, state tomography reveals the accumulated phase $\phi(x,t)=\tfrac{\mu_B g_J}{\hbar}\Delta B(x)\cdot t+\phi_0$, where $\phi_0$ is a constant phase accumulated during the shuttling. By performing such measurements for different phase accumulation times $t$ at different locations $x$, we map out the qubit frequency shift across segments 18-22 with a mean accuracy of about $2\pi\times$1~Hz. With the positions $x_i(t)$ computed from the sequence data and simulated electrostatic trap potentials \cite{SINGER2010}, the phases $\phi_i$ can be also computed and used for correcting the phases of the analysis pulses.
For each of the 16 prepared states, 9 measurements are performed. Each measurement is independently repeated 1000 times. For each prepared state, a resulting density matrix is obtained via linear inversion. From these density matrices, the process $\chi$-matrix is obtained via a second linear inversion.  Computing the trace norm $Tr\left(\chi_{meas}^{\dagger}\chi_{ideal}\right)$, we find a mean process fidelity of 98.1(5)\%. We also perform process tomography without a SWAP operation, obtaining a mean process fidelity of 98.7(4)\%. Thus, on the given level of accuracy, we conclude that the SWAP operation does not significantly affect the measured process fidelity, which is limited  mainly by readout errors and systematic errors of the correction phases. Applying correction for readout errors, we obtain a mean process fidelity of 99.5(5)\%. The resulting $\chi$-matrix is displayed in Fig. \ref{fig:tomo2}.

%%%%%%%%%%%%%%%%%%%%%%%%%%%%%%%%%%%%%%%%%%%%%%%%%%%%%%%%%%%%%%%%%%%%
%\section{Three ion experiments}
%%%%%%%%%%%%%%%%%%%%%%%%%%%%%%%%%%%%%%%%%%%%%%%%%%%%%%%%%%%%%%%%%%%%

\begin{figure}[h!tp]\begin{center}
\includegraphics[width=0.35\textwidth]{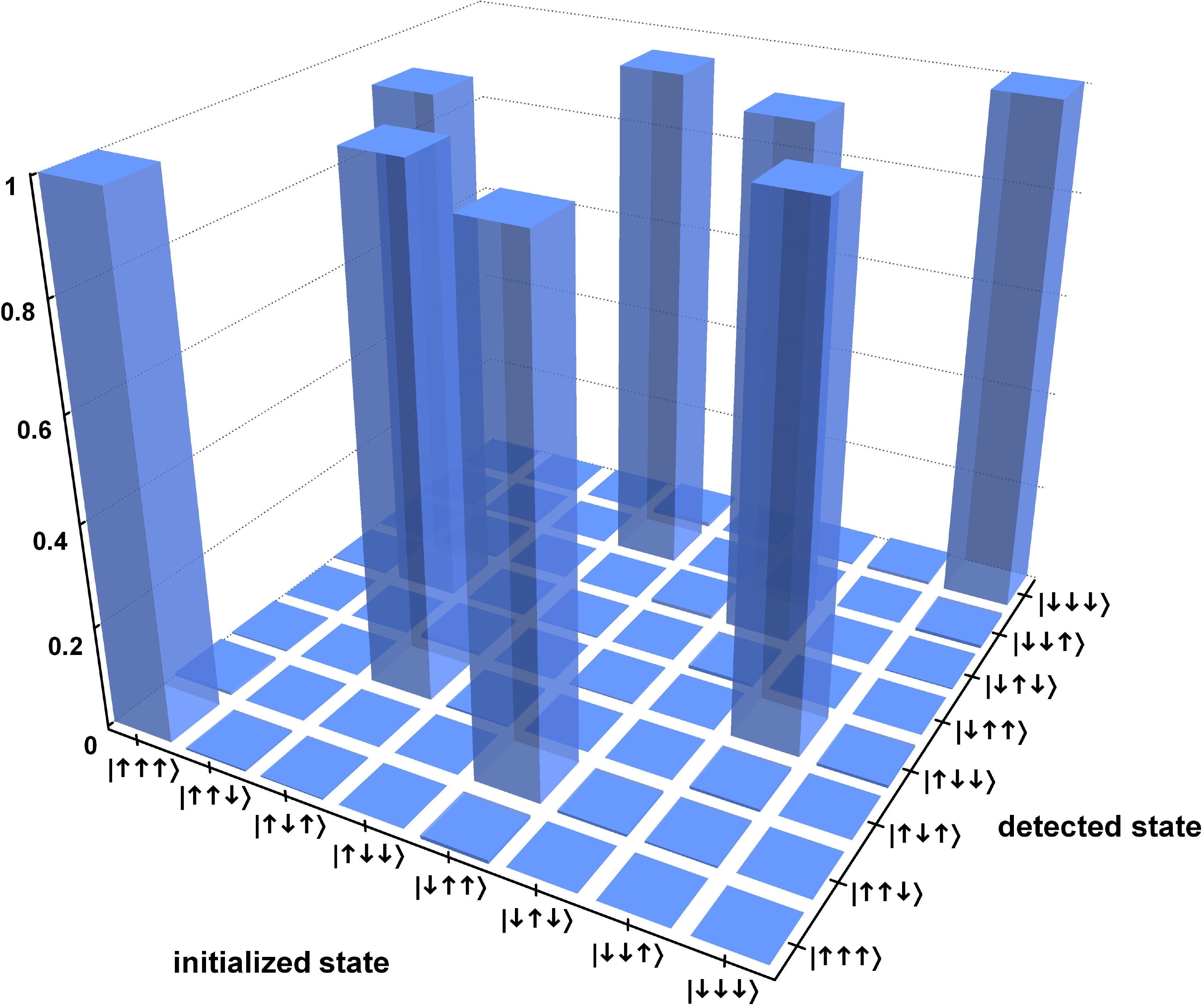}
\caption{Measured truth table of a three-ion crystal reconfiguration from $ABC$ to $CBA$ by using three consecutive two-ion SWAP operations. Each ion was initially prepared in either $\ket{\uparrow}$ or $\ket{\downarrow}$, thus eight different input states are tested. Each input state is prepared and probed on average 2500 times. The measured probability to detect a particular state is represented by the height of the bars.
}
 \label{fig:tomo3}
\end{center}\end{figure}

The techniques described above are extended to three qubits, where we demonstrate reordering from configuration $ABC$ to configuration $CBA$. The detailed sequence can be found in the supplemental material \cite{SUPPLEMENTAL}. Rather than performing quantum process tomography, we restrict the measurements to the logical ($Z$) basis, thus we reconstruct the logical truth table of the reordering operation. Starting from a three-ion crystal $ABC$ in the LIZ, we separate into $AB$ and $C$ by performing the separation with a properly adjusted axial bias field. Ion $C$ is moved to segment 26, then ions $AB$ are moved back into the LIZ, where separation into $A$ and $B$ takes place. Then, $A$, $B$ and $C$ are subsequently moved into the LIZ and initialized to either $\ket{\uparrow}$ or $\ket{\downarrow}$ by optical pumping. Now, the ions are merged pairwise at the LIZ, where swapping and subsequent separation take place. The respective third ion is stored six segments away to the left or right, such that its trapping potential does not affect the merging, swapping and separation operations. The three subsequent swappings $AB\rightarrow BA$, $AC \rightarrow CA$ and $BC \rightarrow  CB$ establish the desired order $CBA$. Then, the ions are individually moved to the LIZ for shelving, and then individually moved again to the LIZ for fluorescence readout. We measure the resulting spin configuration for eight different input states. The resulting truth table is shown in Fig. \ref{fig:tomo3}. We obtain a mean fidelity of 98.47(9)\% in the logical basis. The mean fidelity with correction for readout errors is 99.96(13)\%. The sequence consists of three separation, three merging, three swapping and 30 linear transport operations. The execution time of this process is 5.7 ms, where 93\% of this time is devoted to shuttling operations.

%%%%%%%%%%%%%%%%%%%%%%%%%%%%%%%%%%%%%%%%%%%%%%%%%%%%%%%%%%%%%%%%%%%%
%\section{Outlook conclusion}
%%%%%%%%%%%%%%%%%%%%%%%%%%%%%%%%%%%%%%%%%%%%%%%%%%%%%%%%%%%%%%%%%%%%

In conclusion, we have demonstrated basic functionality of a quantum processing unit based on different shuttling operations, including qubit register reconfiguration. It is shown that operations such as initialization, coherent manipulation and readout are not affected by swapping and other shuttling operations. In future experiments, the time required for such shuttling operations will be substantially reduced by several measures: Compensation of filter-induced waveform distortion will allow for faster ion motion. A novel waveform generator with a voltage range of $\pm$40~V will enable tighter radial trapping, which will enable swap operations at larger radial trap frequencies. Furthermore, control techniques \cite{FUERST2014,PALMERO2015} may be applied to enable faster shuttling. The shuttling-based quantum information processor requires the execution of entangling gates, which among all operations exhibit the strongest sensitivity to motional excitation. Here, we will exploit the fact that shuttling along the trap axis mainly affects axial modes of vibration, and carry out the entangling gates mediated by radial modes. 

%%%%%%%%%%%%%%%%%%%%%%%%%%%%%%%%%%%%%%%%%%%%%%%%%%%%%%%%%%%%%%%%%%%%
%\section{Acknowledgements}
%%%%%%%%%%%%%%%%%%%%%%%%%%%%%%%%%%%%%%%%%%%%%%%%%%%%%%%%%%%%%%%%%%%%

The research is based upon work supported by the Office of the Director of National Intelligence (ODNI), Intelligence Advanced Research Projects Activity (IARPA), via the U.S. Army Research Office grants W911NF-10-1-0284 and W911NF-16-1-0070. The views and conclusions contained herein are those of the authors and should not be interpreted as necessarily representing the official policies or endorsements, either expressed or implied, of the ODNI, IARPA, or the U.S. Government. The U.S. Government is authorized to reproduce and distribute reprints for Governmental purposes notwithstanding any copyright annotation thereon. Any opinions, findings, and conclusions or recommendations expressed in this material are those of the author(s) and do not necessarily reflect the view of the U.S. Army Research Office.

\bibliographystyle{apsrev4-1}
\bibliography{lit}
\clearpage

\newpage
\begin{figure*}[h!tp]\begin{center}
\includegraphics[width=\textwidth,page=1]{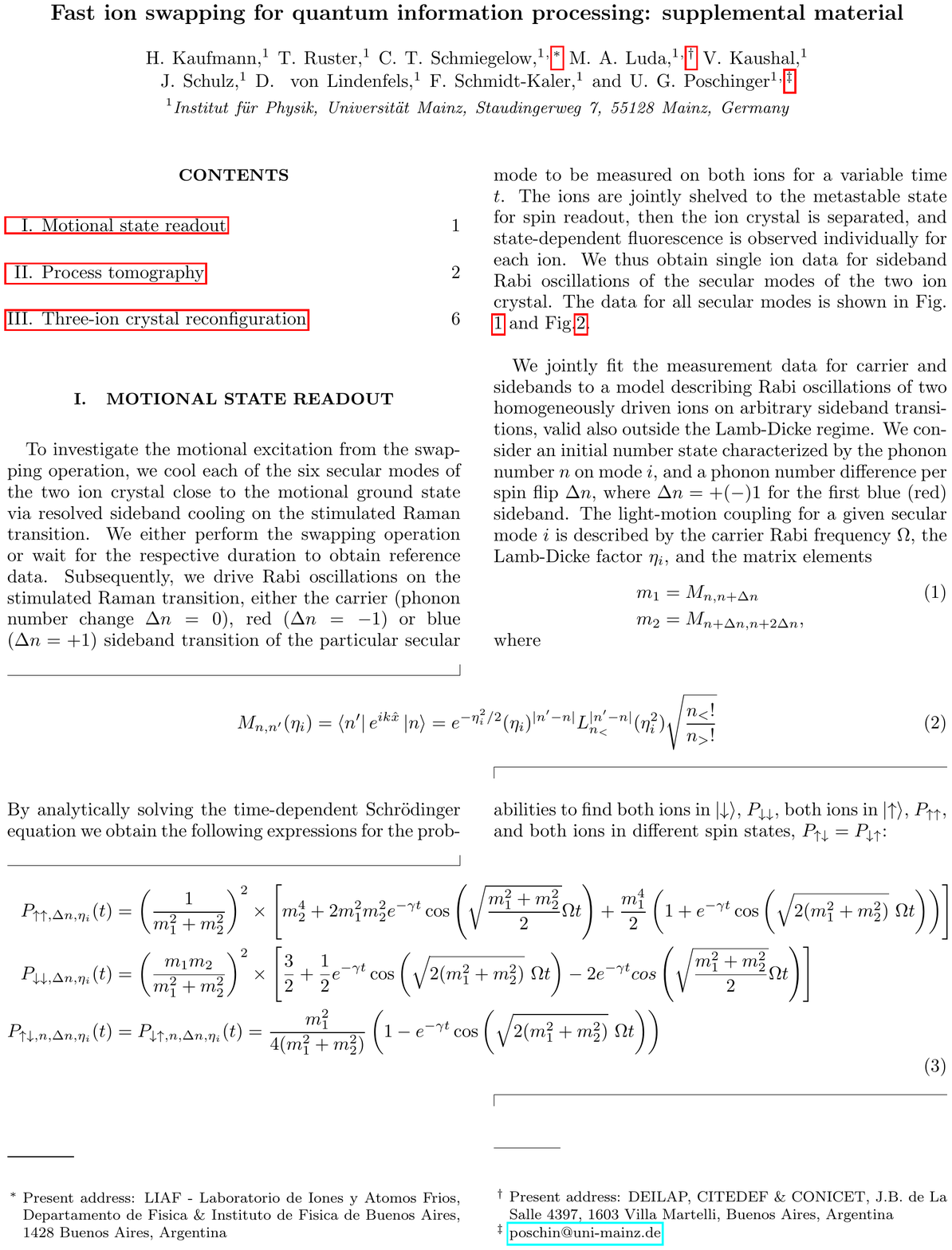}
\end{center}\end{figure*}
\newpage
\begin{figure*}[h!tp]\begin{center}
\includegraphics[width=\textwidth,page=2]{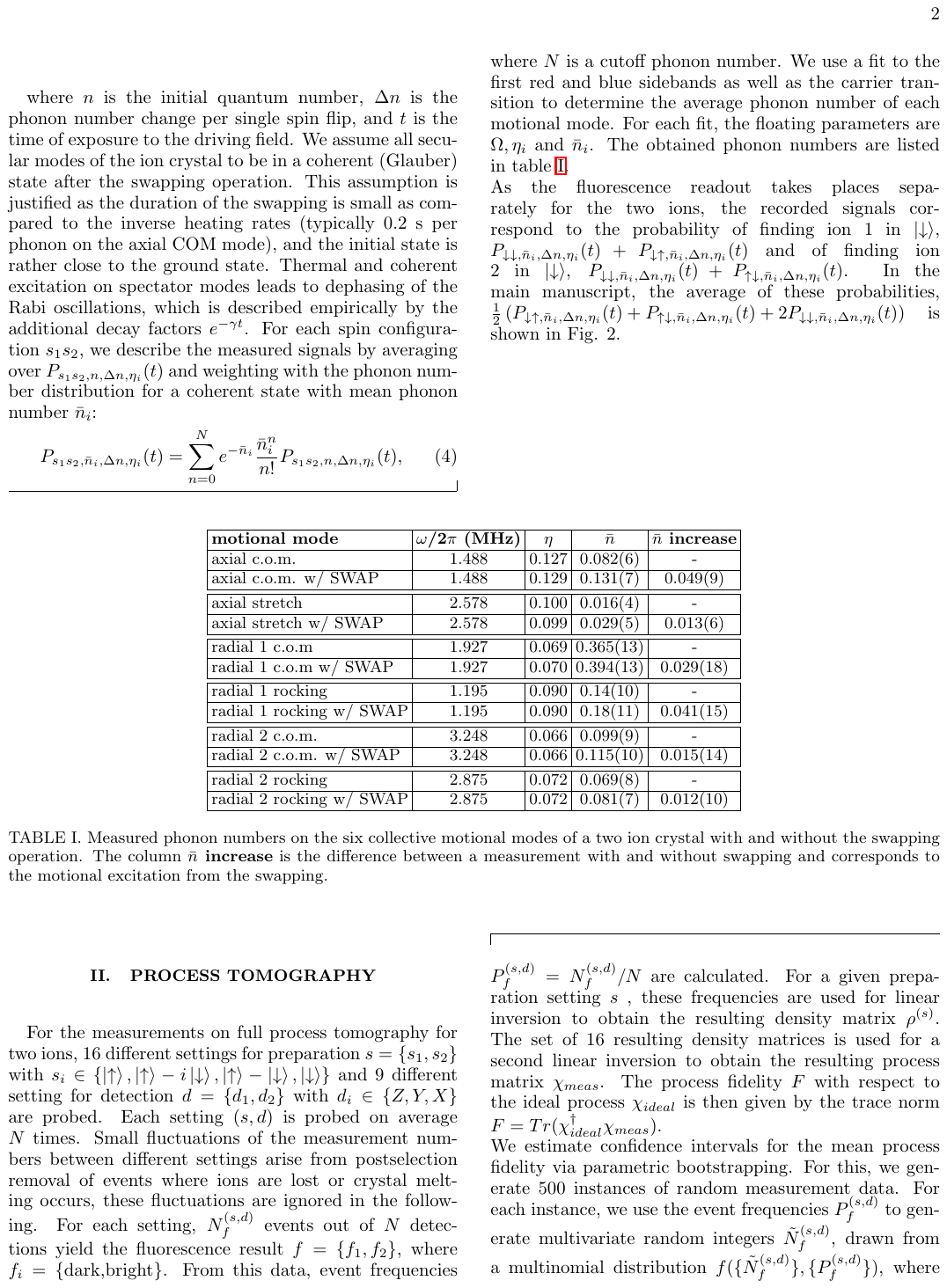}
\end{center}\end{figure*}
\newpage
\begin{figure*}[h!tp]\begin{center}
\includegraphics[width=\textwidth,page=3]{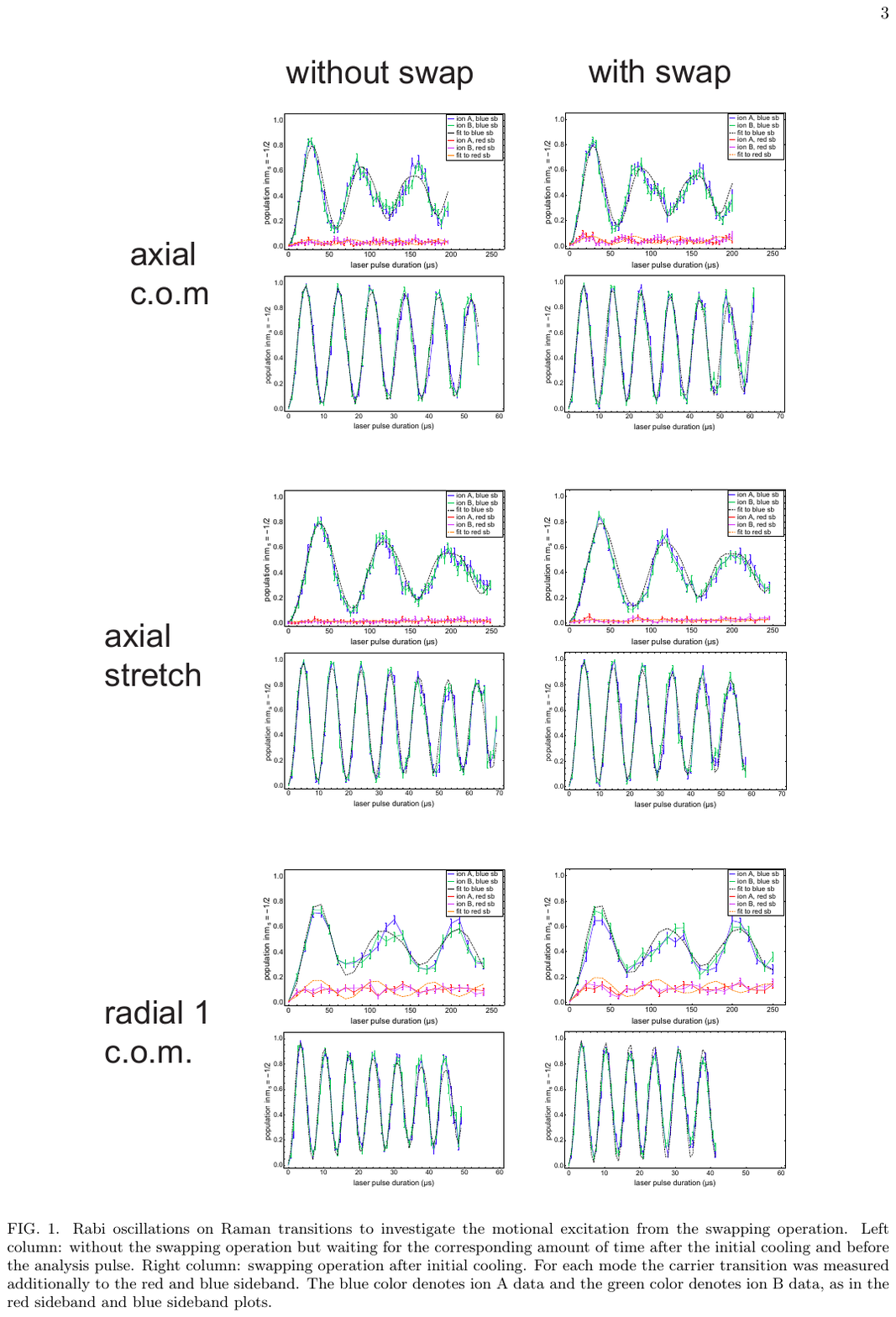}
\end{center}\end{figure*}
\newpage
\begin{figure*}[h!tp]\begin{center}
\includegraphics[width=\textwidth,page=4]{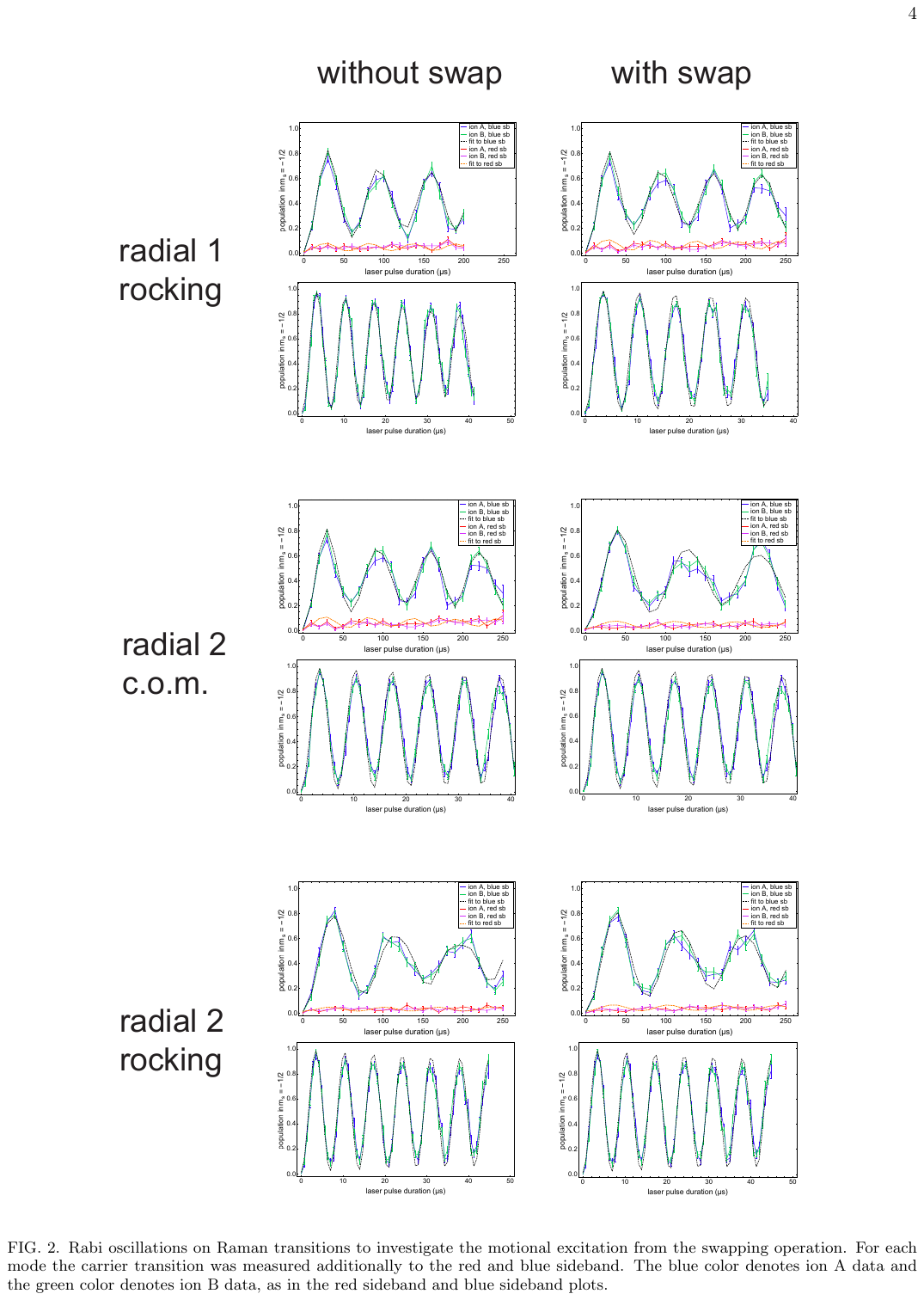}
\end{center}\end{figure*}
\newpage
\begin{figure*}[h!tp]\begin{center}
\includegraphics[width=\textwidth,page=5]{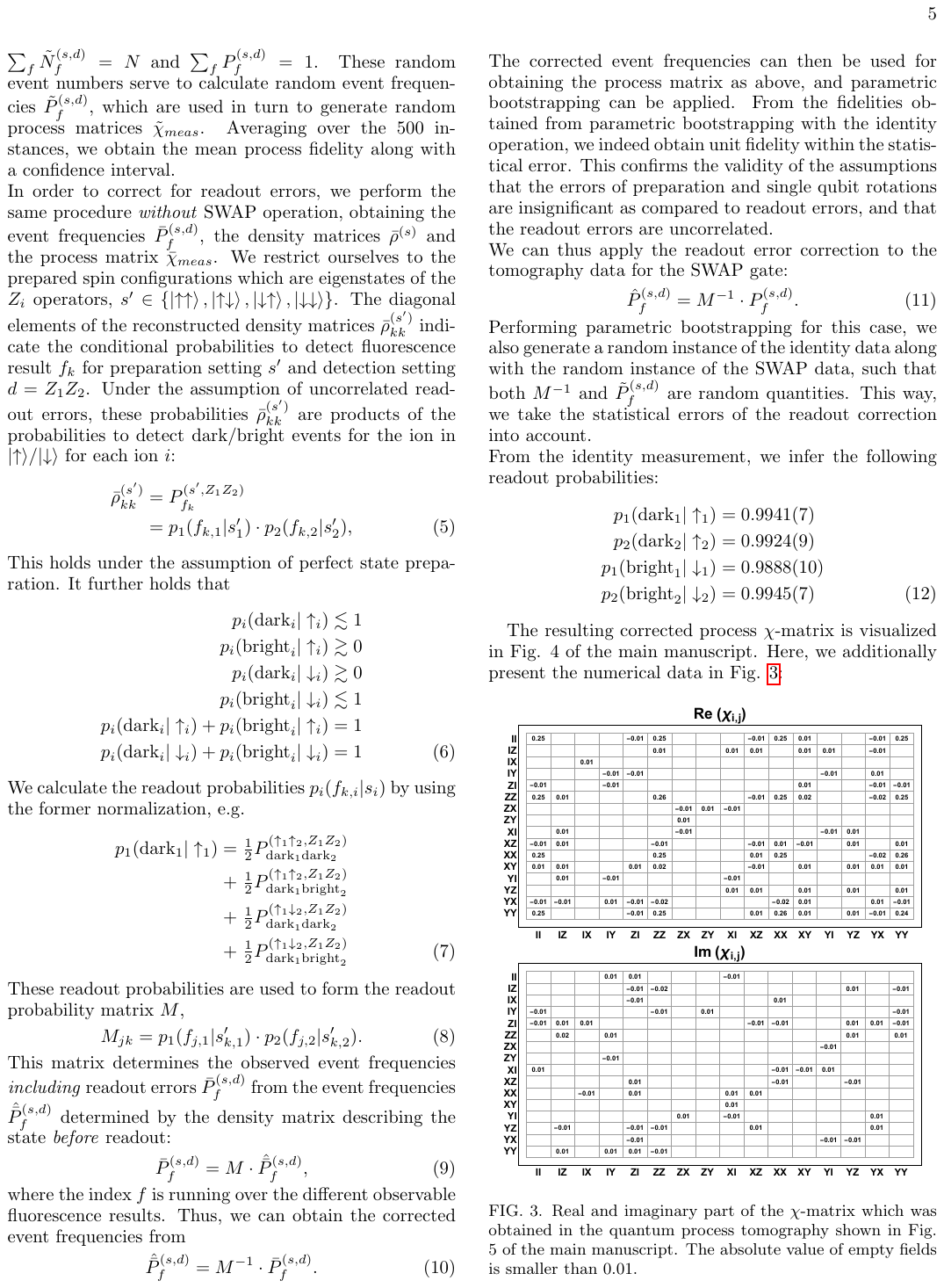}
\end{center}\end{figure*}
\newpage
\begin{figure*}[h!tp]\begin{center}
\includegraphics[width=\textwidth,page=6]{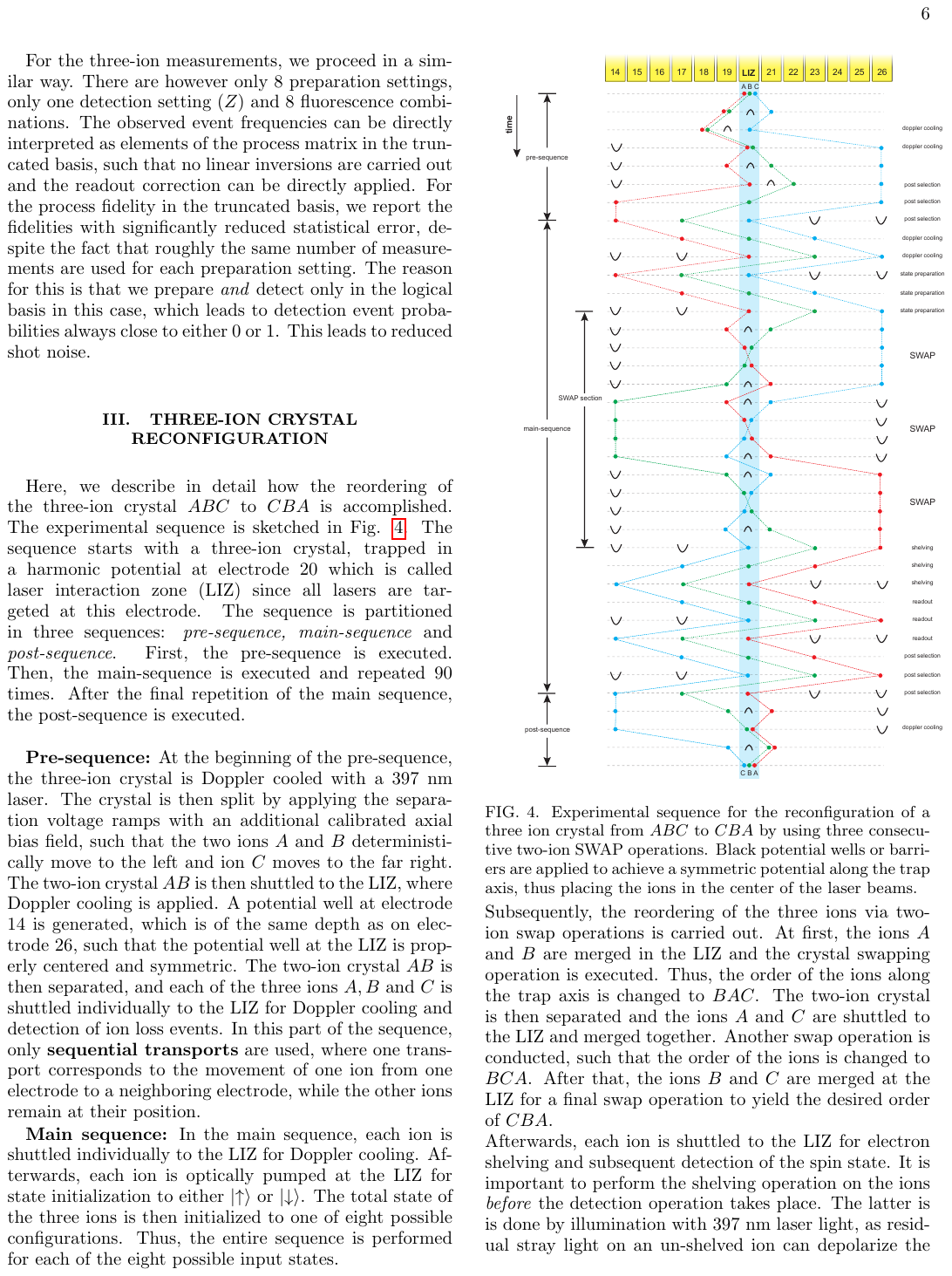}
\end{center}\end{figure*}
\newpage
\begin{figure*}[h!tp]\begin{center}
\includegraphics[width=\textwidth,page=7]{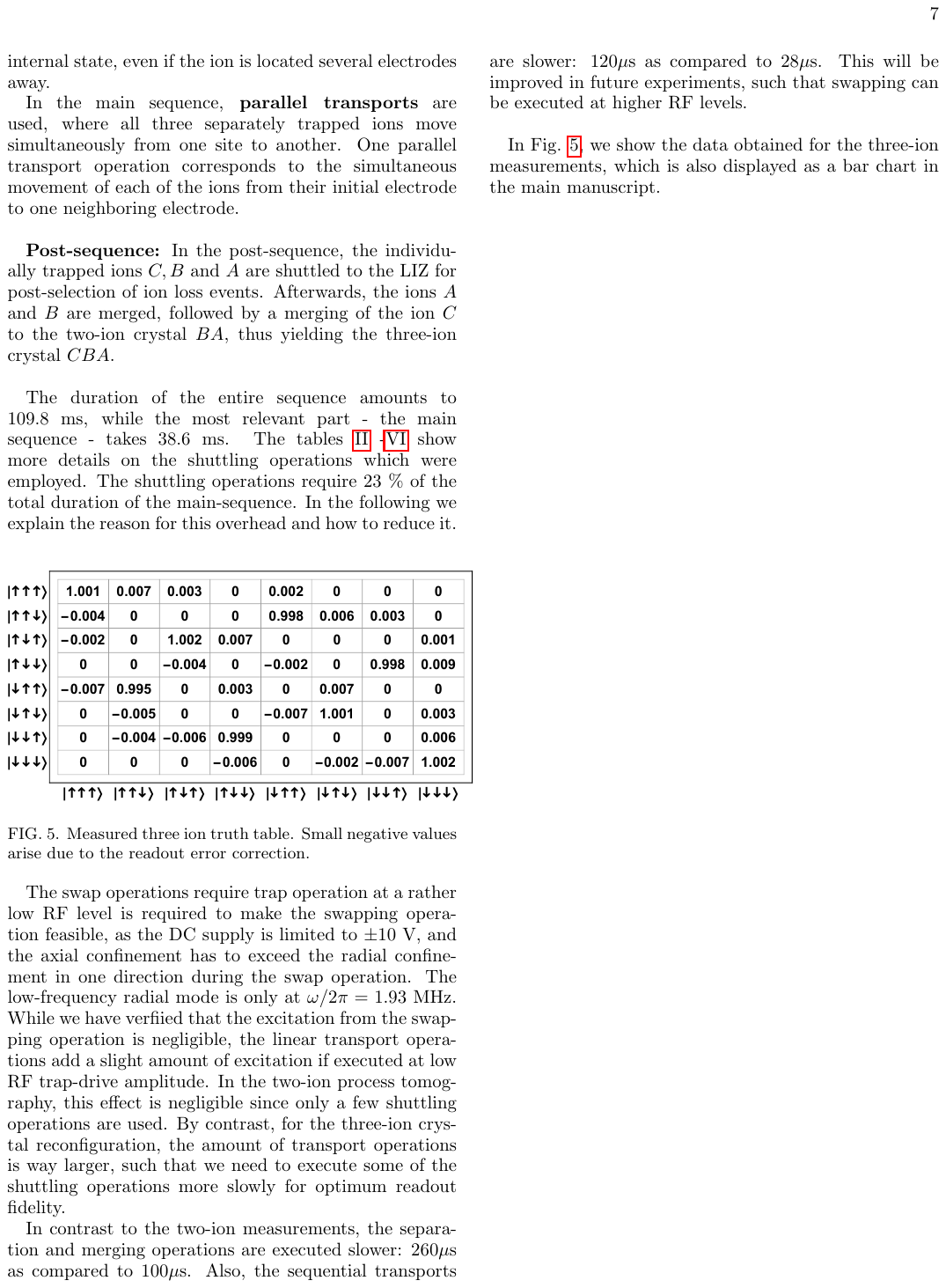}
\end{center}\end{figure*}
\newpage
\begin{figure*}[h!tp]\begin{center}
\includegraphics[width=\textwidth,page=8]{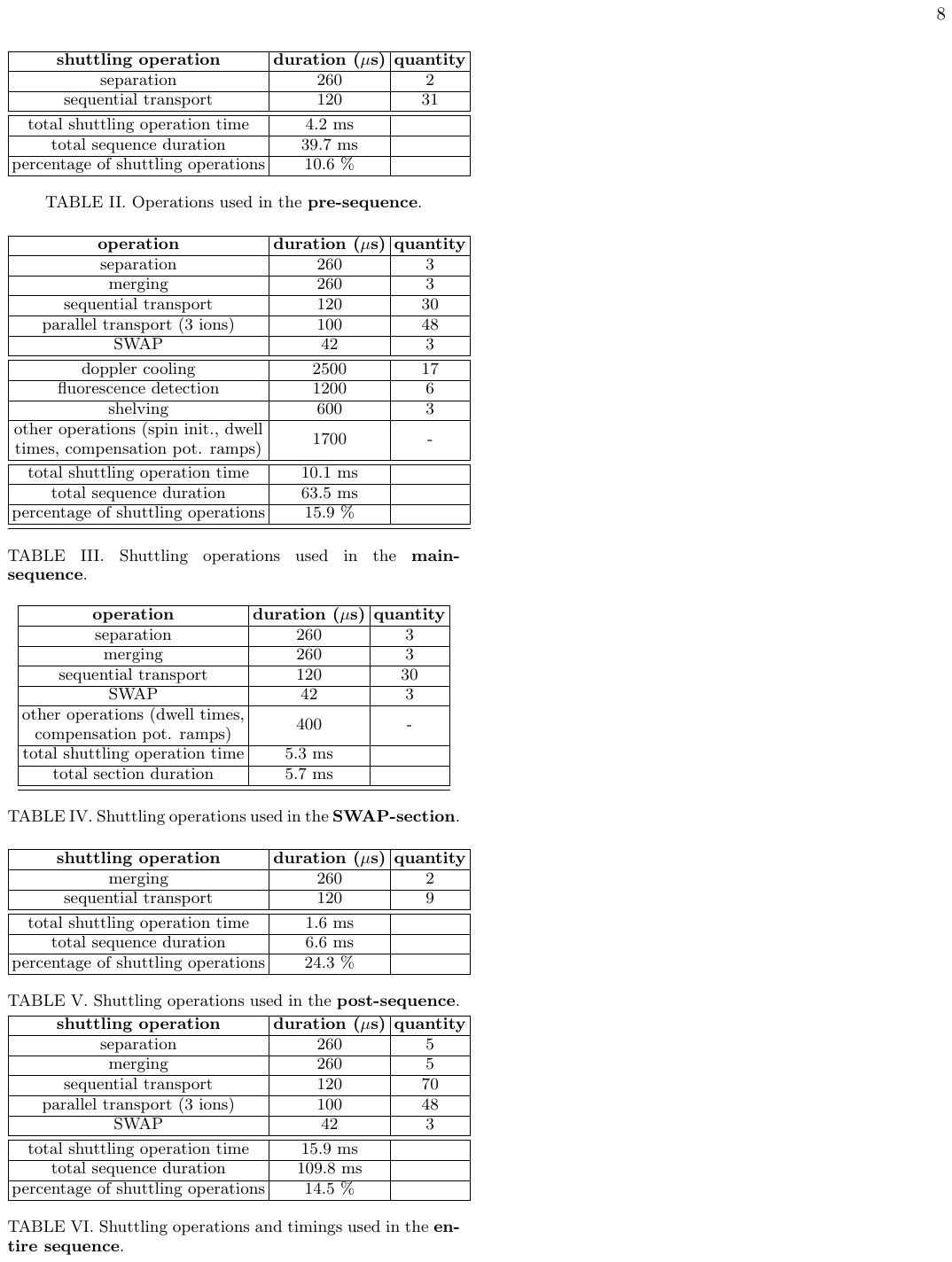}
\end{center}\end{figure*}

\end{document}